\documentclass[aps,prb,twocolumn,floatfix]{revtex4-1}
\usepackage{graphicx,epsfig,array}
\begin{document}

\title{A  Real Space Description of Magnetic Field Induced Melting in the
Charge Ordered Manganites:~ I. The Clean Limit}

\author{Anamitra Mukherjee$^{1}$ and Pinaki Majumdar$^2$}
\affiliation{$^1$Department of Physics and Astronomy, University of British
Columbia, Vancouver, BC, Canada, V6T 1Z1}
\affiliation{$^2$Harish-Chandra  Research Institute,
 Chhatnag Road, Jhusi, Allahabad 211019, India}

\date{\today}

\begin{abstract}
We study the melting of charge order in the half doped manganites using a model that incorporates double exchange, antiferromagnetic superexchange, and Jahn-Teller coupling between electrons and phonons. We primarily use a real space Monte Carlo technique to study the phase diagram in terms of applied field $(h)$ and temperature $(T)$, exploring the melting of charge order with increasing $h$ and its recovery on decreasing $h$. We observe 
hysteresis in this response, and discover that the `field melted' high conductance state can be spatially inhomogeneous even without extrinsic disorder. The hysteretic response plays out in the background of field driven equilibrium phase separation. Our results, exploring $h$, $T$, and the electronic parameter space, are backed up by analysis of simpler limiting cases
and a Landau framework for the field response. This paper focuses on our results in the `clean' systems, a companion paper studies the effect of cation disorder on the melting phenomena.
\end{abstract}

\maketitle
 
\section{Introduction}
Study of correlated materials such as the cuprates, the pnictides and the manganites is motivated as much by fascinating phenomena as by the opportunity to understand many body physics.  In this respect manganites provide a perfect example, hosting phenomena such as colossal magnetoresistance (CMR)\cite{cmr-first, cmr-sec} and multiferroic behavior\cite{multif-first} on the one hand, while offering a window on how strongly coupled degrees of freedom organize  and respond to external stimuli  \cite{mang-book, dagotto1, Zhang201419, Basov2011471}.

This has given great impetus to experimental and theoretical research on phenomenology and microscopic understanding of the manganites.
Experimental work has probed in detail, the doping and temperature dependence of phases and transport properties\cite{mang-book,tok-rev} in the past. More recent focus has been towards controlling phases and transport properties using a variety of schemes such substrate induced strain on thin films\cite{strain-2,strain-3,strain-4, PhysRevLett.109.157203, PhysRevLett.109.157207,Ward:2009be}, heterojunctioning different manganites\cite{Monkman:2012ec,May:2009fd,Hwang:2012hz,Nature_Materials_Gibert:2012kq} and using electric fields\cite{PhysRevB.88.024415,current-2}.  Finally the field of possible device applications has been very active\cite{Hatano:2013hx,device-p,device-2,device-3}. 

Calculations based on model Hamiltonians using  a variety of techniques has culminated in the understanding of the microscopic basis for CMR\cite{dmft,edmc,dagotto1}. Interplay of disorder, strain and impurity physics in the background on phase competition has also been extensively studied, these are discussed later in the text.

There has also been growing interest in investigating impact of time dependent probes such as pump-probe experiments and optical excitation \cite{dyn-1, dyn-2, dyn-3,Nature_Materials_Polli:2007di, PhysRevB.80.115128} on the manganites. These and similar studies with magnetic and thermal field cycling\cite {kuwahara-1,kuwahara-2,cheong,parisi-1,parisi-2,parisi-3,parisi-4} have made understanding of nonequilibrium properties of the manganites very pertinent. At present, however, the theoretical study of nonequilibrium response in the manganites has received little attention. The goal of the present paper and its companion is to address some of these issues. For this we study the effects of magnetic field sweeps on spin charge orbital ordered phases in the half doped manganites. To set the stage, we briefly summarize the basic properties of the manganites.

Among the plethora of phases that the manganites exhibit\cite{mang-book,tok-rev}, of particular interest are the ferromagnetic metal (FM-M) and the antiferromagnetic charge ordered insulating (AF-CO-I) states. The ferromagnetic metal usually shows up in manganites with large bandwidth (BW), {\it i.e}, large mean cation radius, $r_A$, for hole doping $x\sim 0.2-0.5$, while the AF-CO-I state is observed in low bandwidth materials at commensurate doping, $x \sim 0.5$. This `half doped' state is especially interesting since it allows systematic study of phase  competition and the role of disorder. The $x=0.5$ state has been extensively probed experimentally \cite{kuwahara-1,kuwahara-2,respaud,tok_melt1,tok_melt2,tok_melt3,other_exp,cheong,trokiner,melt-exp1a,melt-exp1b, melt-exp2a,melt-exp2b} and also analyzed theoretically \cite{satpathy,fratini,cep_hrk1,other_theory,dagotto1}. 

It is known that low bandwidth materials with  CE magnetic order, checkerboard charge order (CO), and concomitant orbital order (OO), are insulating. We will call this state CE-CO-I. 
The large BW materials have a FM-M ground state. At intermediate BW some materials have `A type' magnetic order. Application of a magnetic field can melt the charge order and convert the CE-CO-I  to a ferromagnetic metal. The  melting transition appears to be
abrupt and is accompanied by hysteresis in response to field sweep \cite{mang-book,tok-rev,kuwahara-1,kuwahara-2}.

Crudely, the CO state in manganites `melts' in response to a magnetic field because the field favors FM order to CE order, and the CO stability depends on the CE order. Some aspects of the melting problem are well studied. (i)~The thermodynamic melting field $h_c$ (which is bracketed by the actual switching fields, $h_c^{\pm}$ , discussed later) is small. The associated energy scale $g\mu_B h_c \ll k_BT_{CO}$, where $T_{CO}$ is the zero field melting temperature ($g$ is the gyromagnetic ratio and $\mu_B$ the Bohr magneton). The smallness of $h_c$ is attributed to the small energy difference between the CE-CO-I and FM-M states.
(ii)~The field induced transition is seen to be first order and is accompanied by hysteresis since there are competing metastable minima, particularly for materials close to the CE-CO-I~-~FM-M phase boundary. (iii)~The field induced FM-M is believed to be  homogeneous, and is so assumed in theoretical studies. These general observations set important benchmarks for any detailed theory, but some key issues remained unresolved.

\begin{figure*}
\vspace{.2cm}
\includegraphics[width=14cm,height=7.2cm,angle=0,clip=true]{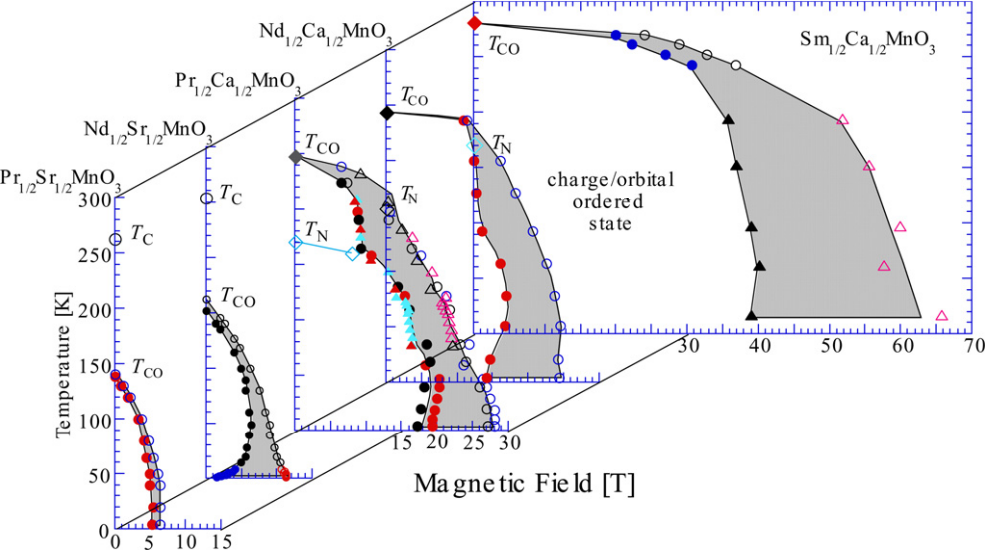}
\vspace{.2cm}
\caption{Colour online: The $h-T$ phase diagram of various RE$_{1/2}$AE$_{1/2}$MnO$_3$ compounds\cite{tok-rev}. The materials involve a systematic decrease in $r_A$ from Pr$_{0.5}$Sr$_{0.5}$MnO$_{3}$ (PSMO) to  Sm$_{0.5}$Ca$_{0.5}$MnO$_{3}$ (SCMO). The critical CO melting temperatures increase with decreasing $r_A$. The associated hysteresis opens a window of metastability at low T, which tapers with increase in temperature and vanishes at $T_{CO}$. The Ca family has low disorder and shows re-entrant behavior in $h_c^{-}$, which vanishes at very low bandwidths (Sm$_{0.5}$Ca$_{0.5}$MnO$_3$). Further, in the Ca family, the decrease in $r_A$, also makes the CO state more robust, with SCMO
having the largest melting fields. The Sr family has larger disorder (see text) with Pr$_{0.5}$Sr$_{0.5}$MnO$_{3}$ having larger $r_A$ than Nd$_{0.5}$Sr$_{0.5}$MnO$_{3}$(NSMO). While NSMO shows marked hysteresis, PSMO is
rather benign. }
\label{f1}
\end{figure*}

(i)~{\it The nature of the finite field state:}
Recent experiments\cite{cheong,trokiner,melt-exp1a,melt-exp1b,melt-exp2a,melt-exp2b}
have demonstrated that the melted state in actually inhomogeneous. We show that
at half doping, for materials with weak charge-orbital order, 
{\it an applied field induces phase separation}
and the `melted' state is at best a percolative metal. 

(ii)~{\it The  field driven equilibrium transition:}
We contend that the {\it equilibrium transition} should be continuous.
Above a threshold,  an
applied field actually leads to phase separation into  AF-CO-I and FM-M regions,
with gradual increase of the FM-M fraction 
with increasing field. The experimentally (and
computationally) observed abrupt field driven 
transition between the CE-CO-I and
the notional `FM-M' states is a 
non-equilibrium effect due to neighboring
metastable states.

(iii)~{\it Switching fields and limits of metastability:} 
While the thermodynamic critical field can be estimated from energy balance, the upper and lower critical fields that are actually measured define limits of metastability, and have not been explored theoretically till now. We calculate these, and compare to experimental scales.

(iv)~{\it The effect of disorder  on field melting:}  
One expects $h_c$ to increase with reducing bandwidth, since the CO is better stabilized. This indeed happens for lanthanides (Ln) of the form \cite{respaud} Ln$_{1/2}$Ca$_{1/2}$MnO$_3$. For members of the Ln$_{1/2}$Sr$_{1/2}$MnO$_3$ family\cite{tok_melt1,tok_melt2,tok_melt3}, with very similar bandwidths, $h_c$ increases initially with decreasing BW but takes a {\it downturn}  beyond a critical BW and then drops to zero, in sharp  contrast to the `divergence' of $h_c$ seen in the 
Ca family. We explain  this in terms of the reduced stiffness of the CO state in the presence of disorder\cite{long-disorder}.

In our earlier short paper \cite{short} we had touched upon some of the issues above. This paper  provides a thorough exploration of the parameter space of the underlying model, focusing on the interplay of equilibrium phase separation and the sweep rate induced non-equilibrium effects.

We have mapped out the $h-T$ phase diagrams capturing both hysteresis and re-entrant features as seen in experiments. In addition to thermodynamic indicators we have characterized the system at low temperature through direct spatial snapshots and by measuring the volume fraction of the CO-I and the FM-M. Our results allow us to provide comprehensive answers to issues (i)-(iv) listed earlier. Further, by putting our results within a Landau-like energy landscape, we have provided a broad framework for organizing the  materials systematics. 

The paper is organized as follows. In section~II  we summarize the key experimental results, and follow it in section~III with a discussion of earlier theoretical work on field melting. In section~IV we define our model and describe the method for solving it. Section~V discusses the zero field reference state. Section~VI is the heart of the paper and discusses the results at finite
field. In sections VII-VIII, we analyze these results in terms of alternate simpler calculations, provide a Landau framework. We conclude in section~IX.

\section{Experimental results}

\begin{table}[h!b!p!]
\begin{tabular}{lllllllll}
\hline
\\[3pt]
Ca & Sr & Ba & & & & & &
\\
\vspace{.3cm}
1.34 & 1.44 & 1.61 & & & & & &\\
La & Pr & Nd & Sm & Eu & Gd & 
Tb & Ho & Y \\
1.36 & 1.29 & 1.27 & 1.24 & 1.23 & 1.21 & 1.20 & 1.18 & 1.18
\\[3pt]
\hline
\end{tabular}
\caption{Ionic radii (in Angstroms) for various AE$^{2+}$ and
RE$^{3+}$ ions in the perovskite manganites\cite{tok-rev}.}
\end{table}

Experiments  on the field response of the CO state have  probed hysteresis, the bandwidth dependence of the melting fields and the effect of disorder through transport measurements and spatial imaging of magnetic correlations in the lattice. We briefly review the important results here.

The key material systematics are embodied in the magnetic field-temperature ($h-T$) phase diagrams for different materials, shown in Fig.\ref{f1}. These are based on the investigations of two lanthanide families, the Ca series \cite{respaud}, Ln$_{0.5}$Ca$_{0.5}$MnO$_3$ , and the Sr series\cite{tok_melt1,tok_melt2,tok_melt3}, Ln$_{0.5}$Sr$_{0.5}$MnO$_3$. The bandwidth is varied by making materials with $Ln$ atoms with progressively smaller radius  (see Table. 1). In 
Fig.\ref{f1}, $r_A$ reduces progressively from Pr$_{0.5}$Sr$_{0.5}$MnO$_{3}$ (PSMO) to 
Sm$_{0.5}$Ca$_{0.5}$MnO$_{3}$ (SCMO).
The phase diagrams were constructed by sweeping up and down in magnetic fields,
at fixed temperature, on samples that had been initially zero field cooled. The melting field in the upward sweep $(h_c^+)$ differs the most from the `recovery' field $(h_c^-)$ in the downward sweep when $T \rightarrow 0$. This difference narrows and vanishes as $T \rightarrow 
T_{CO}$. 

$h_c^-$ shows  a `re-entrant' feature in the intermediate BW materials, {\it decreasing with reducing temperature}. Reducing BW (or reducing $r_A$) progressively increases the stability of the CO state, with SCMO having the largest $h_c^{\pm}$ and $T_{CO}$. 

The half-doped materials are also, inevitably, disordered. The Ln and alkaline earth atoms usually have different ionic radius, and an `alloy', with these randomly located lead to variations in the local electronic parameters. For example, it leads to random changes in the
Mn-O-Mn bond angles,  $\theta$ say, modulating the local hopping,  $\propto cos^{2}(\theta)$. The other effect is `charge scattering' since the Ln and alkaline earth have different valence.
Typically the extent of structural disorder is quantified by the variance $\sigma_A$ of the ionic radii of the A site ions  and its effect has been studied in the Ca, Sr, and Ba families 
\cite{atfld1,atfld2,tomioka}. 
Near $x=0.5$, the variance in the Ca family is $\sigma_A \sim 10^{-3}$\text{\AA}$^2$, for the Sr family greater mismatch leads to $\sigma_A \sim 10^{-2}$\text{\AA}$^2$. We will consider
the Ca family to represent the a `clean' manganite, the Sr family is typical of moderate disorder, while the Ba family involves strong disorder.

The impact of disorder on the zero field $x=0.5$ state has been beautifully demonstrated \cite{akahoshi} by careful preparation of 'non-disordered' samples. This experiment focused on the transition temperatures as a function of disorder. More recent experiments have begun to explore the spatial nature of the melting process \cite{cheong,trokiner} and the role of
disorder \cite{melt-exp1a,melt-exp1b,melt-exp2a,melt-exp2a} in it. The following broad picture has emerged from these studies: 
(a)~The melting fields increase with decreasing $r_A$ in the Ca family but they are {\it suppressed}  on decreasing $r_A$ in the Sr family. For the Ba family, long range CO is absent even at zero field due to the large structural disorder. (b)~Spatial probes suggest, for example, that in  NSMO \cite{trokiner} the  low $T$ finite field state is inhomogeneous, with `poor FM' domains coexisting with perfect FM regions. Similar results were reported in
La$_{\frac{5}{8}-y}$Pr$_{y}$Ca$_{\frac{3}{8}}MnO_{3}$ where CO-I regions are shown to coexist with FM-M regions. (c)~Some experiments \cite{melt-exp2a,melt-exp2b} in LCMO, PCMO, PSMO and NSMO at $x=0.5$ indicate coexistence, at low $T$,  of competing AF-I and FM-M phases with $h-T$ protocol dependent tunable volume fractions.  Additionally, short temporal  magnetic field pulses on (LCMO at  $x=0.5$)  has been reported to cause a switching effects in the volume fraction of charge order, precisely anticorrelated to the magnetic pulse\cite{parisi-3}. 

\section{Status of theory}

Theory of manganites is fairly evolved. Manganites have been modeled with varying degree of realism with a number of techniques including dynamical mean field theory(DMFT)\cite{dmft}, density functional theory+DMFT\cite{PhysRevLett.107.197202}, variational approaches\cite{fratini,cep_hrk1}, and exact diagonalization coupled with  Classical Monte Carlo\cite{dagotto1,edmc}. These have been used to study low temperature phases\cite{dag-1-phases,maj-1-phases,brink-1-half-dop-phases,brink-2-novel-phases,brey-1-phases,brey-2-phases,brink-1-zp,brink-2-zp}, doping and disorder effects\cite{short,maj-1-dis,maj-2-dis,maj-3-dis}, dynamical properties\cite{brink-1-orb-dyn}, transport\cite{dag-1-trans,maj-1-trans,maj-1-dis} and the CMR response\cite{dag-1-cmr,dag-2-cmr,dag-3-cmr}. Beyond bulk manganites, manganite heterojuctions\cite{dag-1-hetero,dag-2-hetero}, strain effects on thin films\cite{,dag-1-trans,brey-1-strain,anam-1-strain}, are being investigated. Further, the coupling between spin, charge and orbitals has been studied to understand multiferroic behavior\cite{dag-0-mulf,dag-2-mulf,dag-3-mulf,brink-1-mulf,brink-2-mulf,brink-3-mulf}. 

However the area of nonequilibrium response to external perturbations has however received little attention. 
Given current experiments probing photo excitation of correlated phases, phase fractions dependence on path taken in temperature-magnetic field variation protocols and field sweeps effects, clear understanding of non equilibrium physics is very relevant. Before we turn to our results, below we briefly survey the limited literature existing in this area.

More specifically at half doping, the metal insulator transitions with changing bandwidth with isovalent A site substitution has lead to remarkable agreement between theory and experiment. The stability of the small BW CE-CO-I phase and the nature of charge order extensively discussed. Indeed the has been considerable debate between Zener polaron type charge order, involving both Mn and oxygen on the one hand and charge disproportionation involving only Mn atoms on the other.

There have been some attempts at a theory of field induced melting in CO manganites
\cite{satpathy,fratini,cep_hrk1,other_theory}. 

(i)~The earliest attempt \cite{satpathy} involved the  mean field study of a one band model with on-site and nearest neighbor Hubbard interaction in addition to double exchange. On application of a magnetic field, the zero field AF-CO-I state was shown to melt to a FM-M through a first order transition. The result shows that a CO state could be destabilized by a magnetic field which couples only to the magnetic sector.

(ii)~A variational study was done for a more realistic model \cite{fratini} incorporating, Jahn-Teller interaction and a class of charge ordered/metallic states with variety of magnetic order (FM, G-AF, CE-AF). This established that decrease in BW resulted in an  increase in magnetic melting (energy crossing) fields.

(iii)~Finally, a two orbital model was studied, with a large family of variational states in a recent work \cite{cep_hrk1}. This established that the smallness of the (thermodynamic) melting field is due to the closeness in energy of the CE-CO-I and the FM-M phases. For a range of electron-phonon coupling they discovered that a CE-CO state with `defects' appears to be lower energy than a pure CE-CO or FM-M when $h \neq 0$. We believe that the result hints at a field induced inhomogeneous state but the authors did not pursue the issue further.

These experimental and theoretical results set the stage for our attempt at understanding the unresolved questions mentioned in the introduction. In particular we (i) study the spatial character of the charge and spin state under magnetic fields,  (ii) examine the `order' of the field melting transition, (iii)~map out the dependence of the switching fields $h_c^{\pm}$ on bandwidth and sweep rate, and (iv)~explore the impact of disorder on the melting process. 

\section{Model and method}

\subsection{Model}

For studying the non disordered problem, we consider a two band model for $e_g$ electrons, Hund's coupled to $t_{2g}$ derived core spins, in a two dimensional square (2D) lattice. The electrons are also coupled to Jahn-Teller phonons, while the core
spins have an AF superexchange coupling between them. These ingredients are all
necessary to obtain a CE-CO-I phase.
\begin{eqnarray}
H &=& -\sum_{\langle ij \rangle \sigma}^{\alpha \beta}
t_{\alpha \beta}^{ij}
 c^{\dagger}_{i \alpha \sigma} c^{~}_{j \beta \sigma} 
 - J_H\sum_i {\bf S}_i.{\mbox {\boldmath $\sigma$}}_i 
+ J\sum_{\langle ij \rangle}
{\bf S}_i.{\bf S}_j \cr
&&  - \lambda \sum_i {\bf Q}_i.{\mbox {\boldmath $\tau$}}_i
+ {K \over 2} \sum_i {\bf Q}_i^2 - \mu N -h\sum_i S_{iz}
\end{eqnarray}
\noindent
Here, $c$ and $c^{\dagger}$ are annihilation and creation operators for
$e_g$ electrons and
$\alpha$, $\beta $ are the two Mn-$e_g$ orbitals
$d_{x^2-y^2}$ and $d_{3z^2-r^2}$, 
labeled $(a)$ and $(b)$ in what follows.
$t_{\alpha \beta}^{ij}$ are hopping amplitudes between
nearest-neighbor sites with the
symmetry dictated form:
$t_{a a}^x= t_{a a}^y \equiv t$,
$t_{b b}^x= t_{b b}^y \equiv t/3 $,
$t_{a b}^x= t_{b a}^x \equiv -t/\sqrt{3} $,
$t_{a b}^y= t_{b a}^y \equiv t/\sqrt{3} $, where
$x$ and $y$ are spatial directions 
The $e_g$ electron spin is 
${\sigma}^{\mu}_i=
\sum_{\sigma \sigma'}^{\alpha} c^{\dagger}_{i\alpha \sigma}
\Gamma^{\mu}_{\sigma \sigma'}
c_{i \alpha \sigma'}$, where the
 $\Gamma$'s are Pauli matrices.
It is coupled to the
$t_{2g}$ spin ${\bf S}_i$ via the Hund's coupling
$J_H$.
$\lambda$ is the coupling between the JT distortion
${\bf Q}_i = (Q_{ix}, Q_{iz})$ and
the orbital pseudospin
${\tau}^{\mu}_i = \sum^{\alpha \beta}_{\sigma}
c^{\dagger}_{i\alpha \sigma}
\Gamma^{\mu}_{\alpha \beta} c_{i\beta \sigma}$, and
$K$ is the lattice 
stiffness, and $h$ the magnetic field. We assume it to be 
in the $z$ direction and coupled only to ${\bf S}_i$.
We set $t=1$,   $K=1$,
and treat the ${\bf Q}_i$ and ${\bf S}_i$ as classical
variables. The magnitude (S=3/2) of the core spin is
absorbed in the coupling constants. 
The chemical potential $\mu$
is adjusted so that the electron density 
remains  $n=1/2$ which is
also $x= 1-n =1/2$.

\subsection{Parameter space}

In the manganites $J_H/t \gg 1$\cite{mang-book, dagotto1}. We choose
$\lambda/t$ and $J/t$ such that the ground state is CE-CO-I at $h=0$, 
but close to a FM-M phase, in accordance to the well established closeness of the energies of these phases in the manganites\cite{dagotto1}. Changing $\lambda$ is equivalent to change in (inverse) BW. BW variation in materials (by changing $r_A$) in mimicked by suitably varying $\lambda$. In reality $J$ also changes with BW variation, but for simplicity, we assume it to be independent and explore only a couple of $J$ values, $J =0.10,~0.12$. The choice is justified later in the text.  
\begin{figure}[t]
\vspace{.2cm}
\centerline{
\psfig{figure=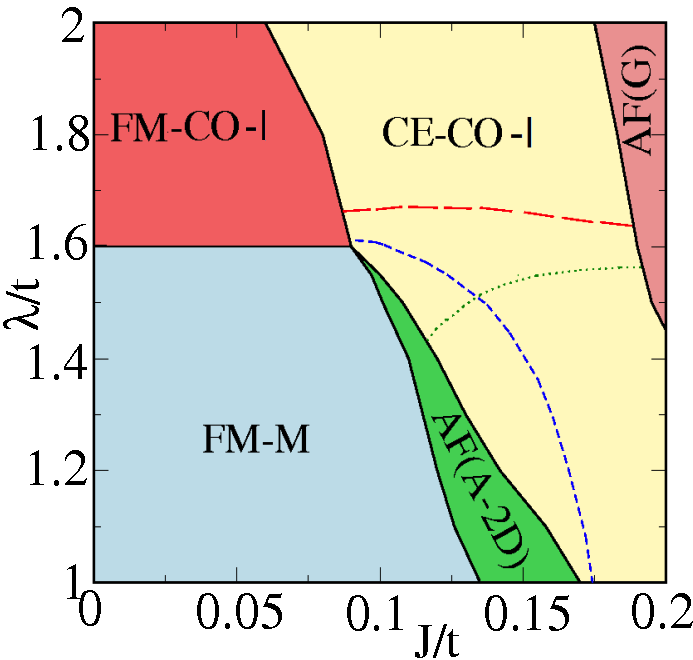,width=6cm,height=5.5cm,angle=0}
\hspace{.2cm}
}
\vspace{.2cm}
\caption{Colour online: The $\lambda-J$ phase diagram at $T/t=0.01$. The
various phases are indicated in the colored regions separated by solid lines. See text for details.  For field induced melting at reasonable $h$, we need to explore the vicinity of $\lambda/t\sim1.6,~J/t\sim 0.10$. The dashed/dotted lines, demarcate various parameter regimes in the CE-CO-I phase in terms of their response to a magnetic field. For large $\lambda$ (above the big-dashed line), the CO state does not
melt on applying a field. Between the big dashed line and the finely dashed line
the CO melts on field application and recovers when the field is reduced.
Between the finely dashed line and the AF-2D phase boundary the field melted CO
does not recover on field removal. Further, the region below the dotted
line, the CE-CO-I state melts to a homogeneous FM-M, while that above it melts to an inhomogeneous state. The zero field phases are discussed in section V and the magnetic response will be discussed in section VI.}
\label{f2}
\end{figure}
\subsection{Method}
We use a variant of the usual real space exact diagonalization (ED) based Monte Carlo (MC) method\cite{dagotto1}. 
In the usual ED+MC the computation cost scales as  $N^{4}$, $N$ being the size of the system. Within our method, the so called  `traveling cluster approximation' (TCA),  the computational cost for the same system is  $\sim NN_{C}^{3}$, where $N_{C}$ is the \textit{fixed} cluster size, and is {\it linear} in $N$ as opposed to $N^{4}$. Since the TCA approach is well established, to avoid repetition, we refer to existing literature\cite{tca} for details. 
Using this technique we have accessed sizes up to $40^{2}$ as opposed to the
limit of $\sim 10^{2}$ within ED+MC. 

The present work is in the spirit of several earlier calculations\cite{madan1,binder1,binder2,binder3}, where
Monte Carlo technique has been employed to explore the non-equilibrium
effects like hysteresis etc., around a phase transition. 
\subsection{Physical quantities}

In order to study the evolution with applied magnetic field, we track various
correlation functions involving the charge, spin, and lattice degrees of
freedom. These include the following:
\begin{enumerate}
\item
The distribution of lattice distortions, $P({Q})=\langle
\sum_i\delta({Q}-{Q_i} ) \rangle$, where $Q_i = \vert {\bf Q}_i \vert$
is the magnitude of the Jahn-Teller distortion at site $i$. Angular brackets
represent thermal average. 
\item
The structure factor for lattice distortions, $D_Q(\textbf{q})=\sum_{ij}
\langle {\bf Q}_i.{\bf Q}_j \rangle e^{i \textbf{q}.({ \bf r}_i- {\bf r}_j)}$. 
This is also a measure of the charge-charge correlation since the local charge
density, $n_i$, approximately follows $Q_i$. 
\item
The magnetic structure factor, 
$S(\textbf{q})=\frac{1}{N^2}\sum_{ij} \langle {\bf S}_i.{\bf S}_j \rangle e^{i \textbf{q}.(
{\bf r} _i- {\bf r}_j)}$. 
\item
The volume fraction of charge order, $V_{CO}$, obtained by analyzing the real
space density distribution. A site with $n_i > 0.5$, surrounded by four sites
with $n_i <0.5$ is part of a CO pattern. Counting such sites allows us to
compute the volume of charge ordered regions {\it even if long range 
order is lost}. This is particularly useful in an inhomogeneous
situation.
\item
The electronic density of states (DOS) is computed as $N(\omega)= \langle {1
\over N} \sum_n \delta(\omega-\epsilon_n) \rangle$, where $\epsilon_n$ are the
electronic eigenvalues in a given equilibrium $\{Q, S\}$ configuration.
\item
The (low frequency) conductivity 
$\sigma_{dc}$, is computed
from the matrix elements of the current operator, as described elsewhere 
\cite{sk-pm-transp}. 
\end{enumerate}
\begin{figure*}[t]
\vspace{.2cm}
\centerline{
\psfig{figure=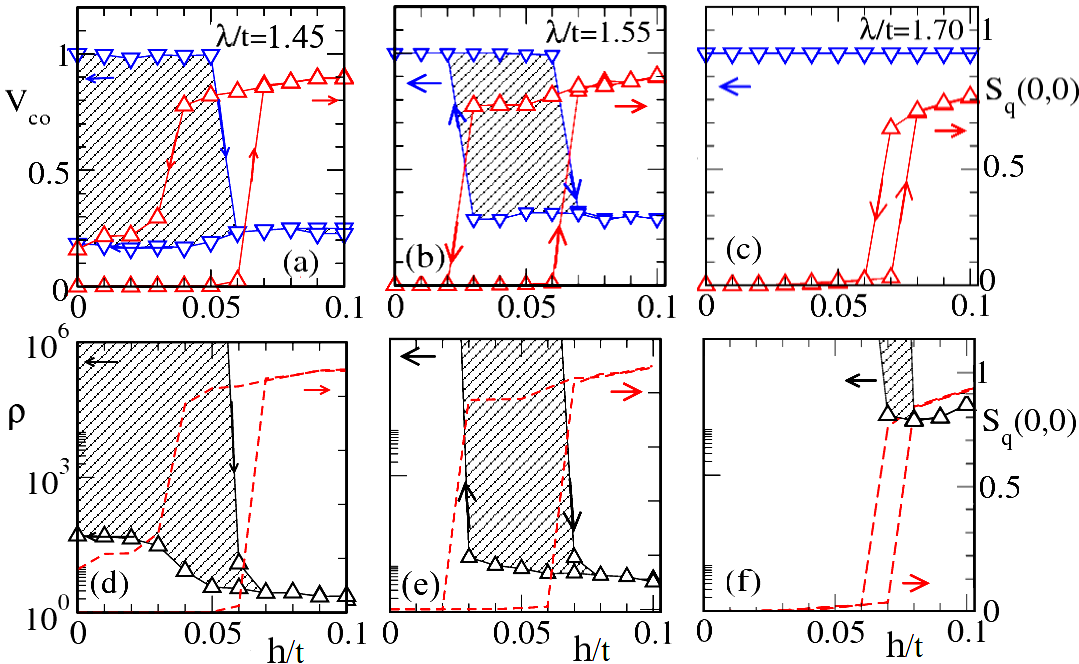,width=13cm,height=8.5cm,angle=0} }
\vspace{.2cm}
\caption
{Colour online: Distinct response to field cycling as a function of $\lambda$.
Top panel, (a) to (c), shows $V_{CO}$ and $S_q(0,0)$ and bottom panel, (d) to
(f), shows the corresponding resistivity as a function of the applied magnetic
field at increasing $\lambda$ values. The $\lambda$ values are indicated, $J/t=0.12$ and T=0.02. In (a) and (b) increasing the field  causes $V_{CO}$ to switch at $h_c^{+}$. This is accompanied by a CE to FM transition as $S_q(0,0)$ shows. There is corresponding drop in $\rho$ as well. In the reverse part of the sweep, $V_{CO}$ switches back to '1'  at  $h_c^{+}$ in (b) but does not recover in (a).  In (c) $S_q(0,0)$ switches between the CE and the FM independent of $V_{CO}$ which remains unresponsive to the field cycling. The FM structure factor is repeated as dashed lines in the lower panels to show the correspondence with resistance switching.}
\label{f3}
\end{figure*}
\section{Results at zero field}

We start with the $\lambda-J$ phase diagram at zero field. It helps identify phases that compete with the CE-CO-I state. This allows us to fix parameter regime for our finite field result. Additionally this phase diagram will be later used to classify different zero field CE-CO-I regimes that  respond differently to magnetic field sweeps. 

Fig.\ref{f2} shows the $\lambda-J$ phase diagram \cite{kp-am-pm1} where
the various phases obtained are indicated. The solid lines are first order phase boundaries  and are determined by annealing from high to low temperature at zero field. The evolution with increasing J/t for $\lambda/t<1.6$ consists of going from FM-M  to A-2D to CE-CO-I to G-AF. The A-2D is a metallic phase with $(\pi, 0)$ or $(0, \pi)$ magnetic order, that lives in narrow region below $\lambda/t\sim 1.6$.
For $\lambda/t<1.6$ the CE-CO-I requires the 'CE' pattern to reduce the BW where upon the electron-phonon (e-p) term becomes dominant causing the CO. However at large $\lambda/t(>1.6)$ we find that the (e-p) term is strong enough to stabilize the CO state even in a FM state as is seen in the top left part of Fig.\ref{f2}. Increasing J/t for $\lambda/t>1.6$ simply evolves the magnetic sector almost independent of the CO state, from FM to CE to G-AF. This implies that for this regime (of small BW) , the magnetic field on the CE-CO-I would cause a transition only to a FM-CO-I and would not melt the CO state. This small BW region is marked by the large dashed (red) line.

We can now justify choice of $J/t\sim0.1)$ values as follows. If we are in a parameter regime where the FM-M and CE-CO-I phases are very close in energy, $\lambda/t \sim 1.6$ and $J/t \sim 0.10$, we can drive a CE-CO-I to FM-M transition by applying a small magnetic field. In the present work we use $J/t=0.10$ and $J/t=0.12$. $J/t=0.10$ allows closer agreement of temperature and magnetic scales to experiments as was shown in our earlier work\cite{kp-am-pm1}. However, $J/t=0.1$ does not allow much room to explore the $\lambda/t$ dependence. The available $\lambda/t$ window  ($\Delta \lambda$) is $\sim 0.1$. At the lower end one hits the A-2D  phase and at the upper end is the large dashed red line, above which the CO-state cannot be field melted. For exploring the BW dependence in more detail we choose $J/t=0.12$ allowing a window $\Delta \lambda \approx 0.2$, while we still remain in the correct ballpark of the experimental transition scales.

The dashed/dotted lines, obtained by magnetic field sweeps at fixed low temperature, indicate the boundaries of the regions showing qualitatively different response to field sweep. These are discussed in the section VI where we study the effects of a finite fields and field sweeps on the CE-CO-I phase.

\section{Results at finite field}

We consider low T magnetic field sweeps to begin with. For understanding the
magnetic field effects we track the various indicators described in Sec.IV.D, by
first cooling the system at $h/t=0$ and then cycling the magnetic field.
In subsection.A we provide a systematic study of the evolution of the field
response with $\lambda$ and in subsection.B we present our $h-T$ phase diagrams and show real space data on inhomogeneous melting.
\begin{figure*}[t]
	\vspace{.2cm}
	\includegraphics[width=17.cm,height=5.0cm,angle=0,clip=true]{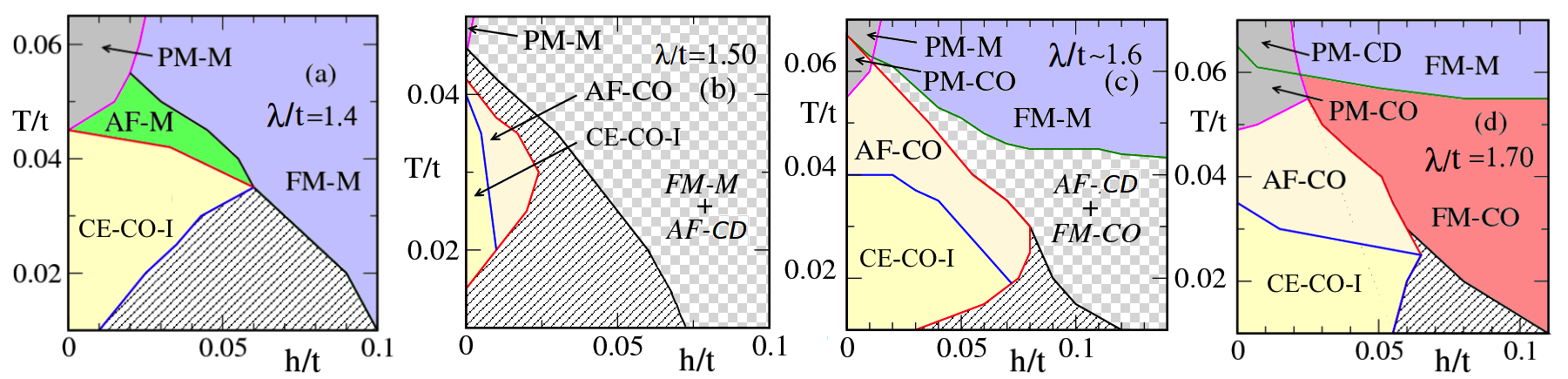}
	\vspace{.2cm}
	\caption{Colour online: 
		The $h-T$ phase diagrams obtained with increasing $\lambda$ values from (a) to
		(d). At any temperature, the shaded regions imply hysteresis, light checkerboard
		regions imply phase separation and colored areas indicate equilibrium phases.
		The composition of the PS states are demarcated in italics.
		The gradual shift of the hysteresis window towards higher fields is
		expected as the CO correlations grow stronger with $\lambda$.  For $\lambda/t=1.4$ and $\lambda/t=1.7$, which are representative of low and high $\lambda$ regimes respectively, the field melted state (beyond $h_c^{+}$) is uniform. For the intermediate $\lambda$ points in (b) and (c), the state beyond $h_c^{+}$  is phase separated. Note that the CO is not recovered for in (b) when the field is swept back to zero at $T/t<0.02$. The various phases in the four panels are discussed in the text.}
	\label{f5}
\end{figure*}
\subsection{Result of a typical field sweep}

Fig.\ref{f3} illustrates the field response for three $\lambda,J$ combinations, with $\lambda$ increasing from left to right. We show the CO volume fraction ($V_{CO}$) in blue and the ferromagnetic structure factor $S_q(0,0)$, in red, in the top panels. The corresponding resistivity $\rho(h,T)$, in black, is shown in the bottom panels. The magnetic field value (while increasing the field) that causes a switching of the CO volume fraction from unity to low values, is denoted by $h^+_c$. Similarly the value of the magnetic field (when sweeping back to zero) that causes the CO volume fraction to switch back to unity, is defined as $h^-_c$ .

Let us consider the broad differences in field response with changing $\lambda$.
(i)~At $\lambda /t= 1.70$, panel 3.(c), the $V_{CO}$ is completely unresponsive
to field change, while the FM structure factor grows and then shows a hysteretic
decrease as expected for antiferromagnetic (CE) to FM transition.
(ii)~At $\lambda/t=1.55$ the CO `melts' in response to increasing $h$,
but only partially, with a residual $V_{CO} \sim 0.3$. Here the magnetic transitions occur concurrently.

The lower panels, Fig.\ref{f3}(e)-(f) show that  $\rho$ remains very large at
all $h$ for $\lambda/t=1.70$, while there is a distinct `switching' for $\lambda/t =
1.55$. This state with a finite $V_{CO}$ and but `low' resistivity at $\lambda /t=1.55$ is likely to be a percolative metal, with CO regions dispersed in a FM-M background. This physics is discussed later. 

For $\lambda/t=1.45$, in Fig.\ref{f3}(a), we again find that the high field ($h>h^+_c$) state $V_{CO}$ is finite, but, remarkably, the CO state is {\it not recovered } when the field is reduced to zero. For even lower $\lambda/t\sim 1.4$ (not shown)
we find a homogeneous FM-M melted state at large fields.

To summarize, we find that depending on response to applied fields as seen in Fig.\ref{f3}, the zero field CE-CO-I region in Fig.\ref{f2}, can be divided into distinct regions. These are demarcated by dashed and dotted lines in the CE-CO-I region in
Fig.\ref{f2}. \\

(i)\textit{Melting vs non-melting:} As discussed in section V and as seen in Fig.\ref{f3}(c), for $\lambda/t\gtrsim1.6$ we find the CO is independent of the CE order. Thus a magnetic field enough to induce a CE to FM transition, would simply push the system in a FM-CO-I phase. For lower $\lambda$, the CO can be melted by a magnetic field. The red (long dashed) line is the boundary.

(ii)~{\it Homogeneous -vs- inhomogeneous melting:} Even when the CO state
responds to a magnetic field, and the long range CO is destroyed, it need not
result in a homogeneous FM state. As seen from the residual V$_{CO}$
for $\lambda/t=1.55$ beyond $h/t=0.07$, there could be phase separation, with a
surviving CO component. Overall, between the large dashed (red) line and the dotted (green) line in Fig.\ref{f2}, the field induced state is inhomogeneous, while below it, the state is a homogeneous FM-M.

(iii)\textit{ Recovery vs non-recovery:} The finely dashed (blue) line separates
regions with different kinds of hysteretic response. Below the red large dashed
line, the CO is recovered in a field sweep if it is above the blue fine-dashed
line. Below this line the CO state is not recovered. 

The issues of metastability in the field response and the equilibrium phase
separation beyond the upper critical field ($h_c^+$) will be addressed in the
next section.
\begin{figure*}[t]
	\vspace{.2cm}
	\includegraphics[width=18cm,height=5cm,angle=0,clip=true]{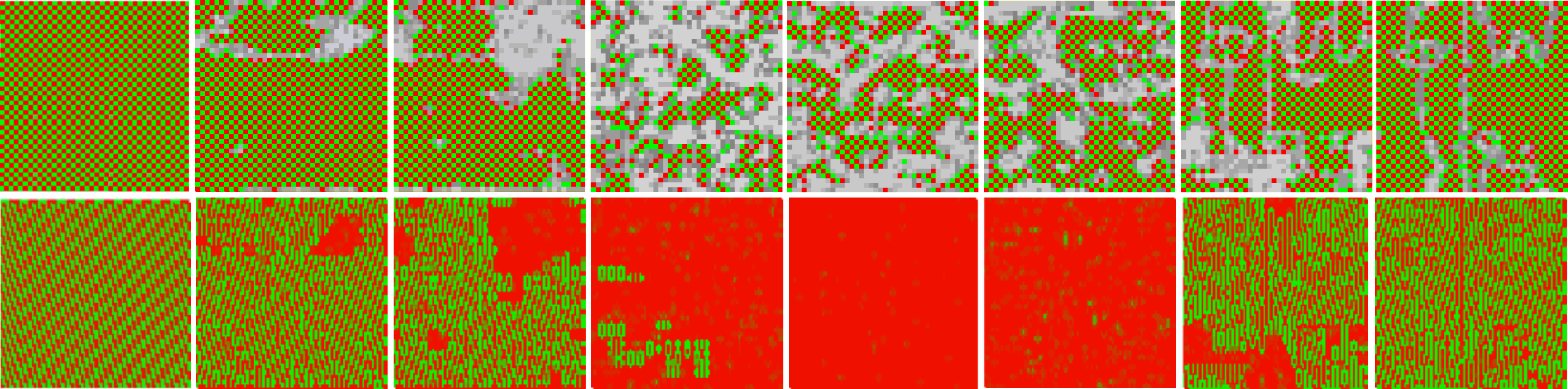}
	\vspace{.2cm}
	\caption{Colour online: The spatial snapshots of charge ordered region (top
		panel) at $\lambda/t=1.55$, $J/t=0.12$ and the corresponding magnetic bonds
		(lower panel) at various magnetic fields during a field sweep. Top panel: red-green checker-board are CO regions and  gray implies metallic (M) regions.  Lower panel,  red are FM bonds, green are AF bonds and the zig-zag pattern are CE regions.
		The field value from left to right are $h/t=~0,~0.05,~0.06,~0.08,~0.20,~0.08,~0.02,~0$ . The CE-CO-I state is almost recovered, as seen in the last column, although for perfect recovery very long annealing is required on this 40$^2$ system.}
	\label{f6}
\end{figure*}
\subsection{The $h-T$ phase diagrams}

In this subsection we discuss the thermal evolution of the low T phases by
constructing $h-T$ phase diagrams. Fig.\ref{f5} shows the $h-T$ phase diagrams obtained at $J/t=0.12$ and $\lambda$ values indicated, increasing from left to right. \\

\textit{Thermal evolution at low fields:} At low $\lambda/t\sim 1.40$ (below the green dotted line in Fig.\ref{f2}), with increase in temperature, the loss of the CE pattern drives the system to an AF-metallic state, with no residual CO correlations. This AF state with S$_q=(\pi,0)$ and S$_q=(0,\pi)$
reflections is a precursor to the low temperature CE phase. This finally leads to a PM-M
at higher temperatures, signifying that the electron-phonon coupling is too weak
to cause an insulating PM state. The hysteresis window expectedly shrinks with
increasing temperature. 
 For $\lambda/t \ge 1.5$ the CO correlations survive at progressively higher temperatures although the long range order is suppressed. These CO regions overlap with the AF regions to form a AF-CO phase, which at higher temperatures give way to paramagnetic (PM)  metal for $\lambda=1.5$  and PM-CO at larger $\lambda$.  Raising the temperature further makes the PM-CO go into PM-M or a charge disorder insulating PM depending on $\lambda$  The details of these phases will be reported elsewhere.
 
\textit{Thermal evolution at large fields:} Let us contrast the low field evolution with that at large finite fields ($h>h_c^+$). Both for small $\lambda$ (1.4) and large  $\lambda$(1.7), the $h > h_c^+$  states are single phases, FM-M and FM-CO respectively. The FM-CO (for $\lambda/t=1.7$) eventually gives way to a FM-M beyond $T\sim0.06$ .

We define the $\lambda$ regime between the red dashed line and green dotted line in Fig.\ref{f2} as the 'intermediate' $\lambda$ regime.

In this regime, for (b) and (c) in Fig.\ref{f5} we find that the $h > h_c^+$ states are inhomogeneous . In the next section we show that the inhomogeneity is due to phase separation (PS) of the system into states of densities different from $0.5$. Moreover the 
constituents vary with changing $\lambda$. At $\lambda/t\sim 1.5$, the coexisting phases are `FM-M ($n_1$) + AF-CD ($n_2$)' and those at $\lambda/t\sim 1.6 $ are 'FM-CO ($n_1$)+AF-CD($n_2$)'. 

The evidence of the phase separation (PS) is also seen in Fig.\ref{f6}, where the top panel shows the spatial charge ordering, while the lower panel shows the corresponding magnetic bonds (see figure caption for color convention) for $\lambda/t=1.55$. These spatial snapshots, at $T/t=0.02$, are for a $40^2$ system and have been obtained from a run in which $h$ is increased from 0 to 0.2 in steps of 0.01, and then reduced to zero in the same sequence. This explicitly shows the inhomogeneous state. However, snapshots based on the Monte Carlo we have employed are likely to be plagued with the system getting stuck in metastable minimas. While we leave the unambiguous determination  of PS to the next section, we conclude this one by mentioning what we observe in the snapshots and by raising a question on the true nature of the melting transition, if indeed these snapshots indicate possible PS.

 In Fig.\ref{f6} beginning with the nucleation of FM-M within the CE-CO-I, there occurs a sharp drop in the CO volume fraction in the fourth column and the system breaks up into an inhomogeneous state. In particular we note that at a very large field of  $h/t=0.2$ (fifth column), the composition of the system is FM-M+FM-CO. Reducing the field recovers to zero the CE-CO-I  state to a large extent (full recovery requires too large a number of states for such big ($40^2$) systems, we have checked that on smaller systems we recover the CE-CO-I phase perfectly). 
 
Assuming the above does points towards phase separation tendency beyond $h^+_C$ raises an important question: Does the equilibrium PS extend only beyond the hysteresis window or does the PS exist at smaller fields as well? If so, how does the hysteresis, between $h^-_C$ and $h^+_C$ occur on the backdrop of an already equilibrium PS state? We answer these questions in the next section.

\section{Nature of melting transition}

Here we deal with two  subtle issues, one relates to 
the existence
of equilibrium phase separation and the second is dependence of the field
response on sweep rates. The second issue is important because typical sweep
experiments are not quasistatic as we infer below and thus it is of interest to
understand the interplay of equilibrium phase separation, hysteresis and sweep
rates.

\subsection{Equilibrium phase separation}

We need to verify that the equilibrium state is indeed phase separated at
intermediate fields. This would be distinct from partial trapping of the system
in some metastable state. We address this via a fixed $\mu$ calculation
described below.

We cool the system at different $\mu$, not necessarily targeting half-filling,
to explore the vicinity of the $x=0.50$ state at finite field. This yields the
$\mu-n$ characteristic, and the various ground states, at finite $h$,
 for a
specific choice of electronic parameters. The $\mu-n $ curves are obtained from
low temperature $\mu$ scans of the system, at fixed $h$, in a protocol that does
not retain the memory of previous $\mu$ steps during the $\mu$ sweep. 

These MC sweeps without memory avoid path dependence, since the system is
annealed {\it ab initio} for each $\mu$, and the fixed $\mu$ character allows 
the system to choose the `best' possible $n$, thereby allowing access to the
correct phase at any $h$. Moreover, we ensured that the system has annealed well
enough by checking that our results hold up to large number (8000) of Monte
Carlo steps at each $\mu$. 
This ensures that the results are well annealed and free from low temperature
Monte Carlo problems. 

As a numerical check we also allow very long relaxation of the phase separated
state within our usual thermal, fixed n, annealing protocol.
\begin{figure}
\vspace{.2cm}
\centerline{
\includegraphics[width=8.5cm,height=4.5cm,angle=0,clip=true]{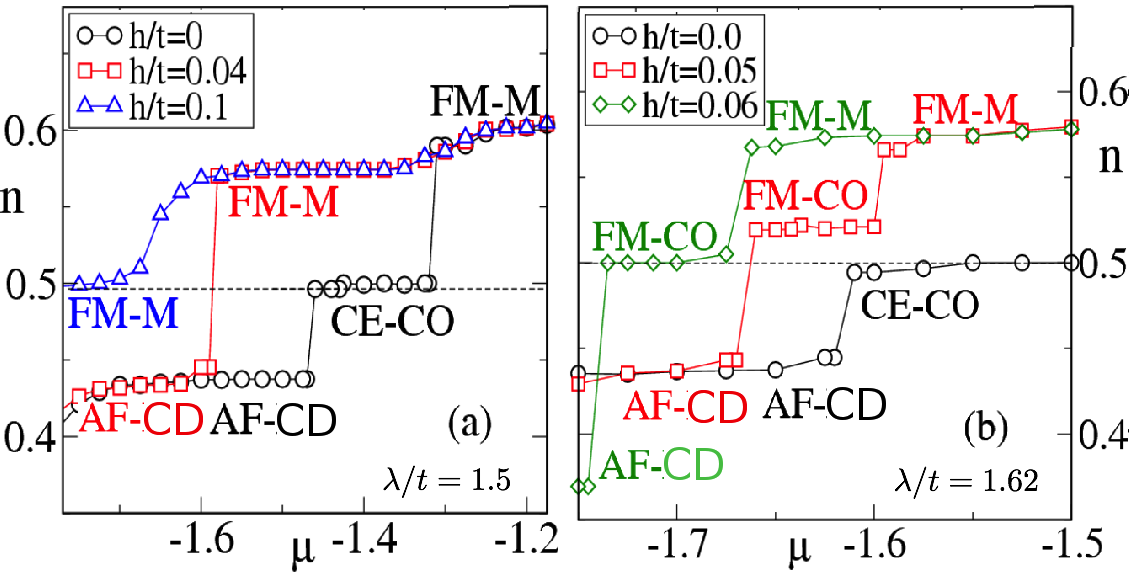}
}
\vspace{.2cm}
\caption{Colour online: $n-\mu$ curves at $T/t=0.02$ for $\lambda/t=1.5$
and $\lambda/t=1.62$ at indicated magnetic fields. Both cases pass through phase separated states, comprising of (AF-CD and FM-M) for $\lambda/t=1.5$ and of (AF-CD and FM-CO) for $\lambda/t=1.62$, before evolving into  uniform FM-M in (a) and FM-CO in (b).}
\label{f7}
\end{figure}
\begin{figure*}[t]
\centerline{
\includegraphics[width=16.5cm,height=4.8cm,angle=0,clip=true]{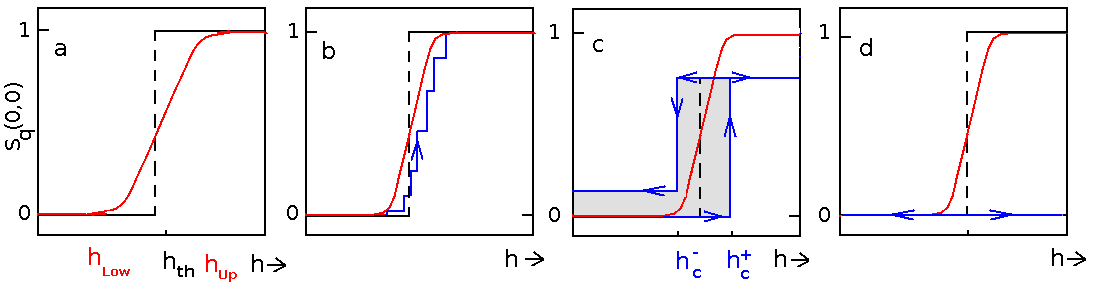}
}
\caption{Schematic sweep rate dependence of the switching at intermediate
couplings ($\lambda$). (a) Equilibrium evolution of the magnetization,
$S_q(0,0)$, with $h$ (red, solid line). $h_{th}$ is a notional value at
which a first order transition would have occurred in the absence of PS. The
evolution of $V_{CO}$ (blue curves) for different sweep rates is shown in
(b)-(d). (b)~Slow sweep, (c)~fast sweep (our regime, see text), and 
(d)~ultrafast sweep. The text discusses the sweep rates in detail. In (c) the
shaded region depicts hysteresis.}
\label{f8}
\end{figure*}

As seen in Fig.\ref{f7}(a), for CE-CO-I systems close to the FM-M phase
($\lambda/t \sim 1.5$) the CO is lost beyond $h\sim0.02$. At a slightly higher
field, the system prefers a  FM-M state with $n=0.57$ up to a certain $\mu$ and
then directly goes to an `A-type' AF  phase at $n=0.44$. If we were to stay at
mean density $n=0.50$ that state would be phase separated, the constituents
being the FM-M and the AF-CD phases. This is true for all systems at
$\lambda/t\sim 1.45-1.6$, and  intermediate $h$. The situation is different for
larger coupling, $\lambda/t\sim1.6-1.65$. For a typical case,
 $\lambda=1.62$ in
Fig.\ref{f7}(b), at intermediate $h$ the system prefers a FM-CO at $n=0.52$
up to a certain $\mu$ and then an AF-CD at $n=0.44$. Again, if we were to stay
at mean density $n=0.50$ the system will phase separate into the above
constituents creating an inhomogeneous state. At larger fields, both
in Fig.\ref{f7}(a) and (b), the $n=0.5$ state becomes stable, recovering the
correct asymptotic limits of FM-M for $\lambda/t=1.5$ and FM-CO for larger
$\lambda/t=1.62$. Apart from confirming the earlier conclusion of inhomogeneous
melting, this calculation helps identify the participants in the PS state. 

If we now consider the spatial snapshots at large fields ($h/t=0.20$), fifth column in Fig.\ref{f6}, we find that coexistence of FM-M and  FM-CO (for $\lambda/t=1.55$).  However from above, we know that the correct ground state at large fields is FM-M for this range of $\lambda$ values.  

This disagreement is however not unexpected, because at  large $h$ (spin polarized limit) the FM-M and FM-CO share a first order boundary and the fixed 'n' Monte Carlo gets stuck partly in the metastable FM-CO minima. This is the reason we performed the calculations presented in this section to determine the true equilibrium PS. Similar calculation for $\lambda=1.4$ and $\lambda=1.7$ do not yield any phase separation.

\subsection{Sweep dependence}
From the $\mu-n$ calculations it is apparent that the melting is inhomogeneous 
for a range of  intermediate e-p couplings. For quasistatic variation of the applied field, for low $\lambda/t$ ($\sim1.4$) and high $\lambda/t$ $(>1.65)$, the transitions are abrupt. For intermediate $\lambda/t$, $\sim 1.45-1.65$, the expected transition is continuous. However as we show here, typical experiments, as  also the magnetic field sweep rate in our MC, do not allow for enough relaxation making both kinds of transitions appear abrupt.

For this we discuss a schematic of such a transition using the ferromagnetic
structure factor as a typical indicator. For intermediate coupling
the magnetic phase separation is between an FM and AF states. From their
densities one can work out the volume fractions of the two constituent magnetic
phases. 

Fig.\ref{f8}(a) shows the magnetization with increasing field. The dashed line
shows
the notional abrupt (first order) transition which is the average of the
critical field for transition in the forward and the backward field sweeps.
The continuous line depicts the {\it expected } `transition' whereby the
magnetization grows continuously (from CE-type AF state) with increasing
$h$ to a FM state. The blue lines are a schematic for the MC
response. The hysteresis that is observed occurs in the background of the
unusual equilibrium physics involving phase separation. Since the magnetization
trace, {\it i.e}, the `switching' in hysteresis,  depends on the sweep rate let
us clarify the experimental and simulation timescales. 

The {\it local} relaxation time $\tau_{loc}$ in electronic systems is $\sim
10^{-12}$ seconds, but collective relaxation times $\tau_{coll}$, say, can be
macroscopic, $\sim 100$ seconds in the CO manganites \cite{sweep-1}.
This experiment was performed at $\sim 0.9 T_{CO}$ and $\tau_{coll}$ is likely to be much greater at low $T$. The field cycling periods $\tau_{per}$ that we could infer
from field melting experiments were \cite{sweep-2}  $\sim 10 $ms. Overall
$\tau_{loc} \ll \tau_{per} \ll \tau_{coll}$. Our MC results are broadly in the
same window. The `microscopic' timescale is the MC step. The sweep periods were
$10^3 - 10^4$ MC steps (bigger in smaller systems) but still $ \ll 10^{12}$ that
one would need to avoid trapping in a metastable state. 

The sweep rate dependence of the switching is illustrated schematically in
Fig.\ref{f8}, for an intermediate coupling system. The left panel, (a),
is for a quasistatic sweep, $\tau_{per} \gg \tau_{coll}$. In this case there
would be only progressive melting and no hysteresis, the system is always in
equilibrium. Panel (b) illustrates the regime $\tau_{per} \sim \tau_{coll}$,
where the sweep rate is still `slow' but the system cannot quite track the
equilibrium state. In this case there could be successive switching. This regime
is also out of computational reach for the system sizes we use. Panel (c) is for
our regime $\tau_{loc} \ll \tau_{per} \ll \tau_{coll}$. The system switches at
$h_c^+$ on field increase, but not necessarily to the underlying equilibrium
state. The magnetization, $V_{CO}$, etc, are determined by the presence of
metastable states. For $h \gg h_c^+$, where the equilibrium state is a
homogeneous FM (at this $\lambda$) the low temperature system can still remain
trapped in the metastable state. We expect similar reasons to cause non-recovery
in the backward sweep when the $\lambda$ and $J/t$ are such that the FM-M is
still metastable when the system is swept back. Finally, (d) is for an ultrafast
sweep, $\tau_{per} \sim\tau_{loc}$, where the system is unable to respond at all
to the changing field.

As shown in panel (c), for sweep rates typical in the experiments and in our
calculation, the high field state is influenced by the equilibrium PS {\it and}
nearby metastable states. This can be seen in the field sweep spatial snapshots of Fig.\ref{f6} for $\lambda/t = 1.55$. At $h/t=0.08$ (column four), we find  that the state arises from a combination of equilibrium AF-M + FM-M
phase coexistence and a metastable FM-CO.  Increasing $h/t$ to 0.2 (column 5)  converts the AF-M to FM-M but the metastable FM-CO fraction can be removed only by thermal annealing. 

\begin{figure*}[t]
\vspace{.2cm}
\centerline{
\includegraphics[width=17cm,height=6cm,angle=0,clip=true]{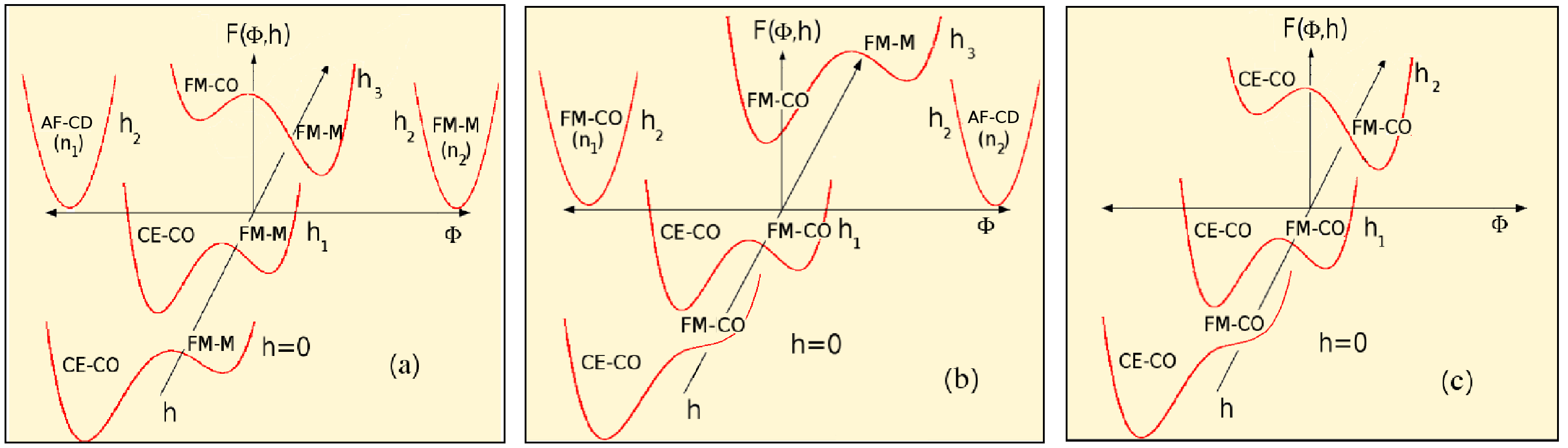}
}
\vspace{.2cm}
\caption{Colour online: The low temperature Landau free energy landscape with
the magnetic field axis going into the plane of the paper and the other
axis, $\Phi$, being the general order parameter axis. The three panels (a), (b)
and (c) are representative of $\lambda/t=1.5$, $\lambda/t=1.6$ and
$\lambda/t>1.65$ respectively. All the phases are at electron density $n=0.5$,
except the ones for which a density are shown in brackets along with the name of
the phase. The intervening A-type AF region is not shown to avoid cluttering.}
\label{f9}
\end{figure*}
\section{Landau framework for field melting}
Over the earlier sections, we drew a number of conclusions regarding the 
$\lambda$ dependence of the magnetic response. Here we suggest a Landau free
energy landscape involving the relevant competing phases and organize the field
response within a single framework. 

While we do not present a Landau functional here, based on our results we
schematically show an energy landscape in terms of some generalized order
parameter. While deriving such a theory from the microscopic model is
difficult, a heuristic construction could still be useful as an organizing
tool. A Landau theory with the provision of stabilizing both commensurate CO at
half doping and incommensurate order off half doping \cite{lit-landau-theory}
and the concomitant magnetic order has been studied before. This reproduces the
qualitative $x-T$ phase diagram around half doping and exhibits phase
coexistence in absence of either strain or disorder. Our landscape can help
improve such constructs. 

From the previous sections we know that the melting can either be homogeneous or
inhomogeneous. For $\lambda/t < 1.6$ the system can melt the CO simply by lowering
the energy of the FM-M minimum  with increase in $h$. The increase in field can
either lead to a simple first order transition, as happens for $\lambda /t\sim
1.40$ or lead to PS as happens for intermediate $\lambda$. In either case the
loss of CO volume fraction is guaranteed. However for $\lambda/t\sim1.6-1.65$, the
FM-CO is closest in energy to the CE-CO-I (and also the true ground state in the
limit of $h\rightarrow\infty$). Without the intermediate $h$ phase separation,
the CE-CO-I would have simply gone over to the FM-CO phase, as happens for
$\lambda/t>1.65$. \textit{The phase separation is necessary for the destabilization of
the CO for this $\lambda$ window}.

With this general understanding, let us discuss the Landau landscape shown in
Fig.\ref{f9}. This has three panels depicting the free energy
landscapes with increasing magnetic fields for three increasing values of
$\lambda$. 

\textit{Small $\lambda$ response: }
Panel (a) of Fig.\ref{f9} corresponds to $\lambda/t\sim1.5$, where the CO
state melts beyond a critical field but does not recover when the field is swept
back. The $h=0$ landscape has CE-CO-I as the global minimum and the FM-M is
metastable. 
This metastable FM-M is responsible for the non recovery of the CE-CO-I state
when $h$ is swept back to zero. From  $h=h_1$ to $h_2$, the FM-M minimum lowers
as expected with increasing field. If the $\lambda/t$ is small, $\sim1.4$, 
this continues leading to a first order transition to a homogeneous FM-M.
However, if $\lambda/t\sim1.5$, at $h_2$ the system phase separates into off half
doping phase (AF-CD + FM-M), these two minima are depicted in panel (a). On
further increasing the field the system evolves into the large field $n=0.50$
FM-M ground state. The phase that is closest in energy to this is the FM-CO, as
is seen in Fig.\ref{f2} at low $J/t$ and $\lambda/t\sim 1.5$. Given the
tendency to get trapped in the FM-CO, we depict this state as metastable at
large fields.

\textit{Intermediate $\lambda$ response:} 
Panel.(b) shows a similar landscape for  $\lambda/t \sim 1.60-1.65$. 
There are a few important differences compared to panel.(a). (i)~From
Fig.\ref{f2} the FM-CO phase is the closest to the CE-CO-I
phase, that can be accessed by a magnetic field. (ii)~Since we know that the
CE-CO phase is recovered when the field is swept back, in the $h=0$ landscape
the FM-CO has to be unstable, as opposed to the FM-M being metastable at $h=0$
in (a). (iii)~The phase separation at intermediate fields is between FM-CO and
AF-CD as depicted, which are off half doping phases. (iv)~Finally, at large $h$
the FM-CO is the global minimum and the FM-M minima is metastable, as is seen 
Fig.\ref{f2} at low $J$.

\textit{Large $\lambda$ response:} This is shown in panel.(c). Like the small
$\lambda$ systems, the large $\lambda$ systems have a simple field  evolution. 
As in (b), the phase closest in energy to the CE-CO is the FM-CO and since the
CE-CO state is recovered when the field is swept back to zero, this FM-CO state
should be unstable at $h=0$. With increasing $h$  the FM-CO energy would lower
and finally replacing the CE-CO as the global minimum. At large fields (not
shown) the CE-CO would become unstable. Note here the CO does not melt and, in
this view, if the intermediate $h$ PS did not occur for $\lambda/t\sim 1.60-1.65$,
CO melting would not have been possible. This we believe is an important
observation.

\section{Conclusions}
We reported the first controlled results on the field melting of charge order in half doped manganites using an unbiased Monte Carlo method. We showed how magnetic field sweep rate induced non-equilibrium physics plays out on the background of equilibrium phase separation, governing the response to magnetic fields and creating inhomogeneous phases without disorder. Our framework to incorporate field melting response within a free energy landscape can aid construction of Landau theories for these materials. 

Recent experiments\cite{step-5, step-4} have followed up older work \cite{step-1,step-2,step-3} on features seen in the magnetization curve with magnetic field sweep. These results are close to half doping or with small doping of the Mn site. The most  recent experiment\cite{step-5} finds step like features for  slow sweep rate (1T/s), which gives way to an abrupt transition with two metamagnetic anomalies for large sweep rates($10^3$T/s). These are consistent with our conclusions discussed in Sec. VII, however the true equilibrium continuous transition can be mapped only for much smaller sweep rates. We hope such experiments would be performed in future.
More generally, these results bring out the importance of relaxation of correlated degrees of freedom in understanding current experiments employing time dependent external probes on such materials.

We acknowledge use of the Beowulf cluster at HRI. PM was supported by a DAE-SRC Outstanding Research Investigator grant, and the DST India.

\bibliography{ref_new.bib}
\end{document}